\documentclass[english]{emulateapj}
\usepackage[T1]{fontenc}
\usepackage[latin1]{inputenc}
\usepackage{amssymb}
\usepackage{amsmath}
\usepackage{graphicx}

\makeatletter

\providecommand{\tabularnewline}{\\}

\usepackage{times}
\usepackage{multirow}

\shorttitle{$\mu$TDEs and ultra-long GRBs}
\shortauthors{Perets et al.}
\makeatother

\usepackage{babel}
\begin{document}

\title{{\normalsize{}{}Micro - tidal disruption events by }\emph{\normalsize{}{}stellar}{\normalsize{}{}
compact objects}\\
 {\normalsize{}{}and the production of ultra-long GRBs}}

\author{Hagai B. Perets\altaffilmark{1}, Zhuo Li\altaffilmark{2}, James
C. Lombardi Jr.\altaffilmark{3} and Stephen R. Milcarek Jr.\altaffilmark{3}}
\altaffiltext{1}{ Physics Department, Technion - Israel Institute
of Technology}\altaffiltext{2}{Department of Astronomy and Kavli
Institute for Astronomy and Astrophysics, Peking University, Beijing,
China}\altaffiltext{3}{Department of Physics, Allegheny College,
Meadville, PA 16335, USA}

\begin{abstract}
We explore full/partial tidal disruption events (TDEs) of stars/planets by \emph{stellar} compact objects (Black holes; BHs; or neutron stars; NSs), which we term micro-TDEs. Disruption of a star/planet with mass $M_{\star}$ may lead to the formation of a debris disk around the BH/NS. Efficient accretion of a fraction $(f_{acc}=0.1$ of the debris may then give rise to bright energetic long ($10^{3}-10^{4}\,s$), X-ray/Gamma-ray flares, with total energies of up to $(f_{acc}/0.1)\times10^{52}\,(M_{\star}/{0.6\,M_{\odot}})\,$ergs, possibly resembling ultra-long GRBs/XRFs. The energy of such flares depends on the poorly constrained accretion processes. Significantly fainter flares might be produced if most of the disk mass is blown away through strong outflows. We suggest three dynamical origins for such disruptions. In the first, a star/planet is tidally disrupted following a close random encounter with a BH/NS in a dense cluster. We estimate the BH (NS) micro-TDE rates from this scenario to be few$\times10^{-6}$ (few$\times10^{-7}$) ${\rm yr}^{-1}$ per Milky-Way galaxy. Another scenario involves the interaction of wide companions due to perturbations by stars in the field, likely producing comparable but lower rates. Finally, a third scenario involves a BH/NS which gain a natal velocity kick at birth, leading to a close encounter with a binary companion and the tidal disruption of that companion. Such events could be associated with a supernova, or even with a preceding GRB/XRF event, and would likely occur hours to days after the prompt explosion; the rates of such events could be larger than those obtained from the other scenarios, depending on the preceding complex binary stellar evolution.
\end{abstract}

\section{Introduction}

The disruption of stars by massive black holes (MBHs) had been extensively
studied over the last few decades, and in particular following the
observational detection of candidate tidal disruption events (TDEs;
see \citealp{Kom15} for a review). However, stars (and sub-stellar
objects such as planets/brown dwarfs) can also be tidally disrupted
by stellar compact objects (COs), such as stellar black holes (BHs),
neutron stars (NSs) and white dwarfs (in this paper we focus on the
former, BHs and NSs; white dwarfs, WDs, will be discussed elsewhere),
with typical masses $10^{-6}$ smaller than the masses of TDE-producing
MBHs. These ``micro''-TDEs ($\mu$TDEs) result from close encounters
between a star and CO. The importance of such close encounters have
been first emphasized in the context of the tidal capture mechanism
suggested by \citet{1975MNRAS.172P..15F} to explain the formation
of close compact binaries (and thus the formation of cataclysmic and
X-ray binaries), and had been later invoked to explain non-standard
formation and evolution of exotic stars such as blue stragglers, Thorne-Zytkow
objects \citep{1977ApJ...212..832T}, and a variety close binary systems
(see \citealp{1999PhR...311..363S} for a short review).

Encounters of COs can be categorized into several possible scenarios:
physical collisions, tidal disruptions, tidal captures and tidal encounters.
These scenarios correspond to the progressively larger distance of
the closest approach of the CO trajectory to the star, respectively.
Here we study the case of tidal disruptions, whereas other cases of
physical collisions or alternatively more distant tidal encounters
of COs with stars have been studied by others \citep{1975MNRAS.172P..15F,1977ApJ...212..832T,1998ApJ...502L...9F,1998ApJ...505L..15H,1999ApJ...516..892F,2001ApJ...549..948A,2001ApJ...550..357Z,2005MNRAS.361..955B}.

Three distance scales are important for describing the encounters:
the closest approach distance $R_{p}$, the radius $R_{\star}$ of
the star, and the tidal disruption radius 
\[
R_{t}\simeq R_{\star}\left(\frac{M_{\bullet}}{M_{\star}}\right)^{1/3},
\]
where $M_{\bullet}$ and $M_{\star}$ are the masses of the compact
object and the stellar (or planetary) object, respectively.

In close (non collisional) encounters the tidal forces can be sufficiently
strong as to completely or partially disrupt the star, in which case
a fraction of the stellar mass may fall back, self interact and eventually
be accreted onto the CO. Although the possibility of tidal disruption
of stars by stellar COs at close encounters was suggested by many,
the observational signature of $\mu$TDEs, their frequency and their
consequences have been little explored. In this paper we discuss this
possibility and suggest that $\mu$TDEs can result in highly energetic
flares, possibly similar to gamma ray bursts (GRBs/X-ray flashes -XRFs),
but much longer (>few $\times10^{3}$ s) and fainter than most of
them. We find that the timescale of these flares is a few tens of
minutes to hours, potentially related to the recently observed class
of ultra-long GRBs \citep{Gre+15,lev+15}. An alternative scenario
in which the debris forms an extended long-lived disk around the compact
object, producing an X-ray source very similar to an X-ray binary,
is not discussed here; we refer to \citet{kro84} for an in-depth
study of this possibility (first discussed by \citealp{1976MNRAS.175P...1H}),
which may occur independently of the early accretion flare on which
we focus in this work.

The paper is organized as follows. We first discuss the properties
of $\mu$TDEs in section 2, and provide a basic estimate of their
rates in section 3. We then discuss the observational implications
for such transient events as well as their remnants and summarize.

\section{The tidal disruption and accretion}

Tidal disruption of stars was discussed in the context of binary formation
through tidal capture \citep{1976ApL....17...87H,1976ApL....17...95H}.
In this context the tidal disruption radius was important as the closest
distance where stars can be captured, but the tidal disruption itself
was only briefly mentioned. Tidal disruption of stars by WDs and NSs
were simulated by \citet{Ruf92}, and \citet{1996JKAS...29...19L},
respectively, but the observational signatures from the following
accretion of debris were not explored. Tidal disruption of a star
by an MBH and its observational signature was discussed by many authors
\citep{1982ApJ...262..120L,1988Natur.333..523R,1989ApJ...346L..13E,1990ApJ...351...38C,1993ApJ...410L..83L,1997ApJ...489..573L,Ulm+98,1999ApJ...519..647K,1999ApJ...514..180U,2000ApJ...545..772A,2004ApJ...610..707B}
and possibly observed in recent years (\citealt{2002ApJ...576..753L,2006astro.ph.12525L,2006ApJ...653L..25G,van+11};
see \citealp{Kom15} for a recent review). We follow a similar analysis
used for this scenario, and complement it using results from hydrodynamical
simulations of $\mu$TDEs. These are used to calculate the relevant
parameters for the disruption of stars by \emph{stellar} COs.

As a star is ripped apart by the tidal forces of a CO, the debris
is thrown in a fan-like fashion into high eccentricity orbits with
a large range of periods, covering a range of specific energy

\[
\Delta E\sim\frac{GM_{\bullet}R_{*}}{R_{p}^{2}}
\]
where $R_{*}$ is the radius of the star and $R_{p}$ is the pericenter
of the orbit \citep{1982ApJ...262..120L}. For cases in which the
star is completely disrupted, simulations show that the mass distribution
of the out-thrown debris is nearly constant as a function of the energy
\citep{1999ApJ...514..180U}. A large fraction of the debris would
later be flung out and become unbound \citep{2000ApJ...545..772A}.
The gravitational energy dissipated in this stage is possibly emitted
as a long and very faint flare, with energetics much smaller than
those expected from the later accretion phase which is the main focus
of this study.

The returning debris streams self-interact and a large fraction of
the bound material becomes unbound, whereas the rest circularizes
and forms a torus (e.g.\ see simulations for a disruption by a WD
by \citealp{Ruf92}) at a radius of about $r_{c}=2R_{{\rm p}}$.
In the next stage, after circularization, the torus formed from the
fallback material is then accreted by the CO \citep{1989ApJ...346L..13E,1994ApJ...422..508K,2000ApJ...545..772A,2002ApJ...576..753L}
possibly producing a flare. In the following we discuss the observational
signature (time scales, energetics) of such flares. We note that \citet{luy+08}
have discussed a related scenario and explored the accretion stage
in a tidal disruption by an intermediate mass BH; they suggest this
scenario leads to jet formation producing a GRB with no associated
supernova.

\subsection{Timescales}

\subsubsection{Fall-back time }

In the following we consider two cases: the case of a low mass ratio
disruption, and the case of a high mass ratio disruption. In the former,
the mass of the disrupted star/planet is assumed to be negligible
compared with the CO mass, and the CO can be assumed to be stationary
at the center of mass of the system. Such a case corresponds to the
tidal disruptions of planets or low mass stars by stellar COs. However,
when the mass of the disrupted object is relatively large (0.1 up
to a few 0.1 of the CO mass), the CO can no longer be assumed to be
stationary or reside at the center of mass of the system. The first,
low mass-ratio scenario had been discussed extensively in the literature
in the context of widely studied TDEs by MBHs; although different
in scale the results should also apply for the low mass-ratio $\mu$TDE
case; we briefly review the results obtained for that case. The high
mass-ratio scenario had been little studied and we therefore run hydrodynamical
simulations of such tidal disruptions to characterize some of their
basic properties.

\paragraph{Low mass-ratio tidal disruptions:}

At the first stage following the disruption, the bound fraction of
the debris, $f_{fall}$, falls back and returns to pericenter. The
first bound material returns after a time

\begin{multline}
t_{min}=\frac{2\pi R_{p}^{3}}{(GM_{\bullet})^{1/2}(2R_{\star})^{3/2}}\\
\approx3.52\times10^{5}\left(\frac{R_{p}}{2.15R_{\odot}}\right)^{3}\left(\frac{R_{\star}}{0.1R_{\odot}}\right)^{-3/2}\left(\frac{M_{\bullet}}{10M_{\odot}}\right)^{-1/2}\,s,\label{eq:disruption_time}
\end{multline}
where the normalization was done for a Jupiter-like planet with radius
$R_{\star}=0.1{\rm R_{\odot}}$ disrupted by a BH with a typical mass
of $10\,M_{\odot}$ at $R_{p}=R_{t}=2.15\,R_{\odot}$ closest approach.
Note that for $R_{p}<R_{t}$ one should replace
$R_{p}$with $R_{t}$ in this Eq. \citep{Sar+10,Gui+13,Sto+13}.
Assuming a flat distribution of debris energies, the late time return
rate of the bound material to pericenter \citep{1988Natur.333..523R,phi89,1999ApJ...514..180U}
is 
\begin{equation}
\dot{M}\sim\frac{1}{3}\frac{M_{\star}}{t_{min}}\left(\frac{t}{t_{min}}\right)^{-5/3},\label{eq:debris_fall_back}
\end{equation}
where the peak return rate occurs at about $t\sim1.5t_{min}$ \citep{1989ApJ...346L..13E}
and half of the fallback debris mass returns by about $t\sim6t_{min}$.
Indeed, such behavior is seen in our hydrodynamic simulations of a
BH tidally disrupting a Jupiter mass planet (see Fig.\ 1). Note,
however, that simulations by \citet{2000ApJ...545..772A} show that
a large fraction of the returned debris later becomes unbound. They
find the total accreted mass of debris to be four times smaller than
found earlier, with an approximately constant accretion rate. Nevertheless,
this does not make a significant change to the overall derived timescale.
We also mention the work by \citet{cou+14} who take a somewhat different
approach and suggest the formation of an extended jet-producing envelope.

\begin{figure*}
\includegraphics[scale=0.4]{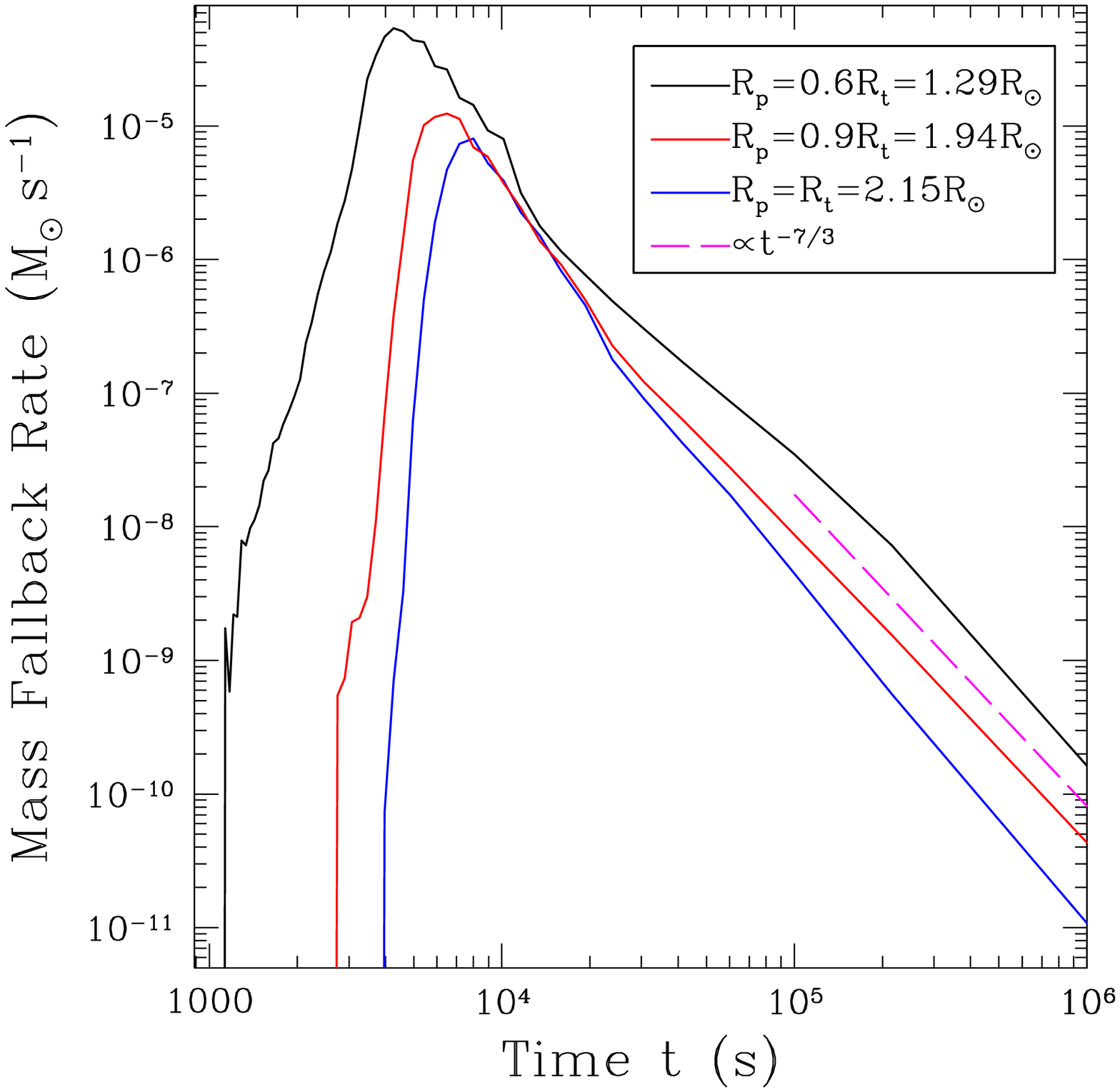}\includegraphics[clip,scale=0.4]{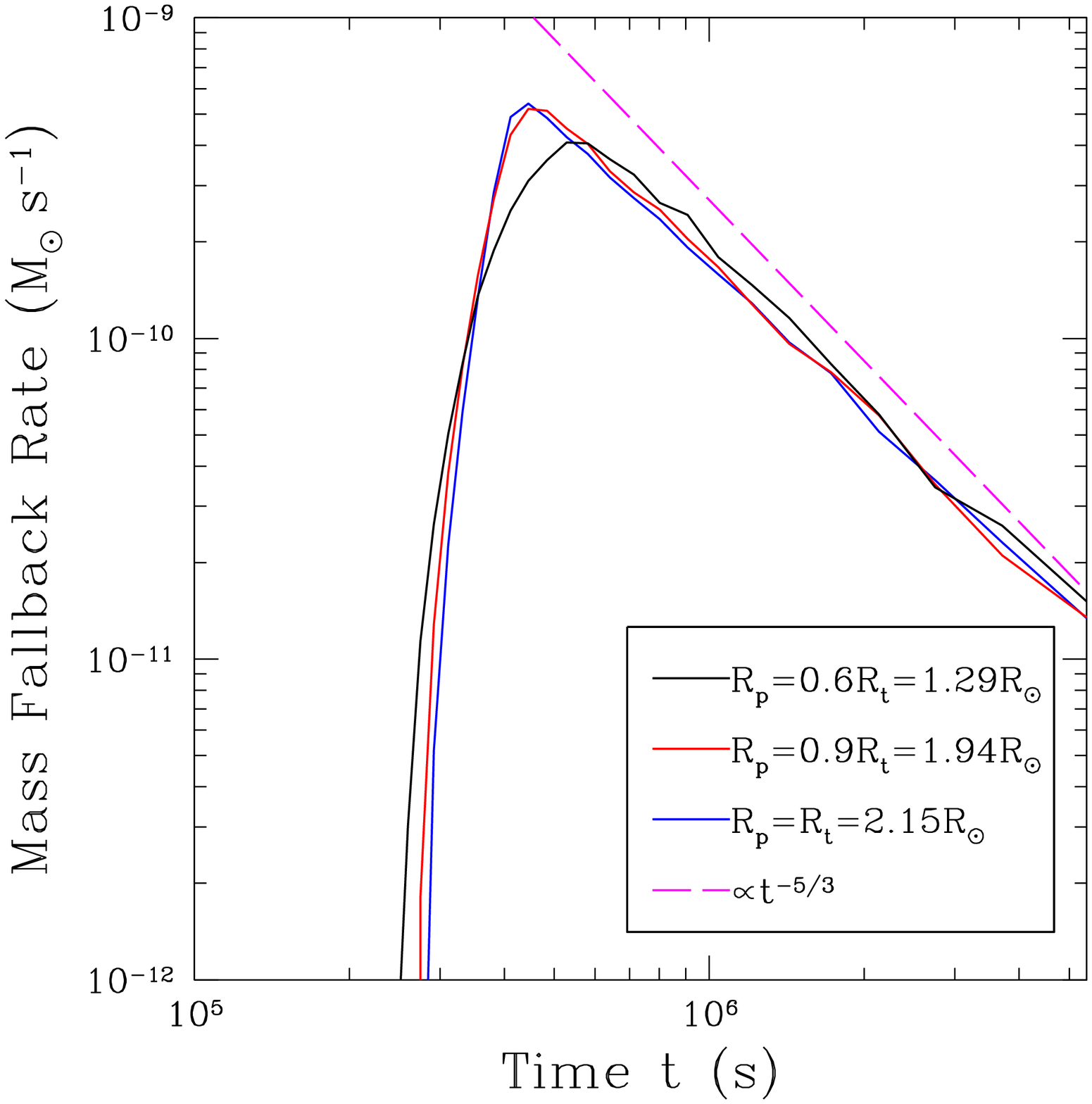}
\caption{\label{fig:The-return-rate}The return rate of debris from a main-sequence
Sun-like star (left panel) and from a Jupiter-like planet (right panel)
due to tidal disruption by a black hole of $10M_{\odot}$. Solid curves
show simulation results for a parabolic orbit with a closest approach
of $0.6R_{{\rm t}}$ (black), $0.9R_{{\rm t}}$ (red), and $1R_{t}$
(blue), where the tidal radius $R_{t}=2.15R_{\odot}$. Dashed magenta
lines show the approximate late time return rates, which is seen to
scale approximately as $t^{-7/3}$ for the partially disrupted Sun-like
star and as $t^{-5/3}$ for the nearly completed disrupted Jupiter-like
planet.}
\end{figure*}

\paragraph{High mass ratio partial tidal disruptions: }

The $t^{-5/3}$ infall rate back to the CO discussed above corresponds
to the case of complete or nearly complete tidal disruption of a low
mass object (compared with the disrupting CO). If the object is completely
disrupted then there is nothing special about the debris field near
gas that is marginally bound to the black hole, and the classic $t^{-5/3}$
behavior follows. The results change somewhat when the object is only
partially disrupted, as can happen more frequently in high mass ratio
cases or when the object has a dense core. The surviving remnant affects
the distribution of gas marginally bound to the black hole, and a
fallback rate steeper than $t^{-5/3}$ results: see e.g.\ \citet{2012ApJ...757..134M}
for simulations of partial disruptions of giant stars by a supermassive
black hole, and discussion of these issues by \citet{Gui+13} and
\citet{Hay+13}.

To study partial tidal disruptions in high mass ratio cases, we have
carried a set of hydrodynamical simulations using the \textit{StarSmasher}
code. \textit{StarSmasher} is a smoothed particle hydrodynamics (SPH)
code that evolved from \textit{StarCrash} \citep{2010ascl.soft10074F},
which itself has its origins in the SPH code of \citet{1991PhDT........11R}.
The primary enhancement of \textit{StarSmasher} over its predecessors
is that it incorporates equations of motion derived from a variational
principle to insure accurate energy and entropy evolution \citep{2010MNRAS.402..105G}.
In addition, gravitational forces between particles are calculated
by direct summation on NVIDIA graphics cards.

We consider the tidal disruption of a solar mass star or a Jupiter-like
planet, modelled with 395K particles, by a 10 M$_{\odot}$ BH point
particle. A wider range of parameter space for tidal disruptions will
be discussed in more detail elsewhere. The solar mass star is evolved
with the TWIN stellar evolution code to an age of 4590 Myr to give
a radius $\sim1R_{\odot}$; the hydrodynamic simulation employ an
equation of state (EOS) consisting of ideal gas and radiation pressure
components. The Jupiter-like planet is modelled as an $n=1$ polytrope
of mass $10^{-3}M_{\odot}$ and radius $0.1R_{\odot}$, with a $\Gamma=2$
EOS. We consider both parabolic and hyperbolic encounters (but we
discuss only the former cases here, more relevant for the expected
relative velocities), and study cases of different closest approach
$R_{{\rm p}}=0.6,\,0.9$, and $1$ $R_{{\rm t}}$ (that is, $R_{{\rm p}}=1.29$
R$_{\odot}$, $1.94$ R$_{\odot}$, and $2.15$ R$_{\odot}$). Runs
with larger closest approach ($R_{p}\sim4$ R$_{\odot})$ resulted
in only a negligible fraction of the stellar (or planetary) mass falling
back onto the BH. None of our runs implement radiative cooling. Although
radiative cooling would be important for generating optical light
curves of the TDEs themselves, it has negligible effect on the hydrodynamics
and therefore on the fallback rates extracted from our simulations.
As can be seen in Table 1, our simulations result in a \emph{partial}
disruption, leaving behind the denser core of the star which is not
disrupted. We did not follow the long term stellar evolution of such
disruption remnants, which could be of interest by itself.

In order to estimate the fallback rate, we use the so-called freezing
model as in \citet{2012ApJ...757..134M}: 
\[
\frac{dM}{dt}=\frac{dM}{dE}\frac{dE}{dt}=\frac{1}{3}\left(2\pi GM_{\bullet}\right)^{2/3}\frac{dM}{dE}t^{-5/3}.
\]
Here Kepler's third law has been used to relate the specific binding
energy $E=GM_{\bullet}/(2a)$, where $a$ is the semimajor axis of
a parcel of debris, to the time $t$ for that material to return to
periapsis. This approximation neglects the influence of the remnant
star, if it exists, on the fallback time. Note that the canonical
$t^{-5/3}$ behavior of $dM/dt$ follows from having a flat $dM/dE$
vs.\ $E$ distribution. More generally, $dM/dE$ depends on the strength
of the tidal interaction as well as the density profile of object
being disrupted. Objects that are completely or nearly disrupted tend
to give relatively flat $dM/dE$ at small $E$ and therefore $dM/dt\sim t^{-5/3}$
at late times. In contrast, objects that are only partially disrupted
yield a fallback rate $dM/dt$ that drops off more quickly with time.

The actual $dM/dE$ distribution from our simulations is determined
by sorting the mass of unbound particles into bins of $E$. For each
unbound SPH particle, we calculate $E$ as the sum of its specific
kinetic energy (relative to the black hole) and its specific gravitational
potential energy due to all mass in the system. For our Jupiter-like
planetary disruptions (with $R_{*}\ll R_{p}\sim R_{t}$), unbound
gas forms a long stream with one end that reaches back to the black
hole only once the planet (or what remains of it) has retreated well
outside of the tidal radius. Since the tidal disruption plays out
fully before the return of the gas to the black hole, it is sufficient
to determine the $dM/dE$ distribution from a single snapshot of the
simulation, taken shortly before the first stripped material returns.

In the disruption of our Sun-like stars (with $R_{*}\sim R_{p}\sim R_{t}$),
stripped material from the star encircles the black hole and collides
with other infalling material before the tidal disruption has fully
completed, and it is consequently not possible to use a single snapshot
to determine the fallback mass rate over all times. We therefore use
a sequence of snapshots when binning particles by their specific energy
in these cases. The first snapshot is taken about $10^{3}$ s after
the periapse passage: all particles that can be identified as being
bound to the black hole are included in this binning, although these
particles may have their bin locations adjusted while we step through
an additional $\sim5$ future snapshots spaced over $\sim5$ dynamical
timescales. If a snapshot shows that new
particles have been stripped from the star, then these particles are
included in the binning. If a previously binned particle has not yet
reached 20\% of its expected fallback time $t$, then it is instead
rebinned according to its specific binding energy $E$ calculated
from the more recent snapshot. In this way, preference is given to
the energetics calculated once a particle has withdrawn from the star
and its trajectory is better described by ballistic motion around
the black hole. By using a short enough time interval over which rebinning
can occur (namely, the 20\% of the fallback time $t$), our procedure
avoids issues associated with other particles that have encircled
the black hole colliding with the infalling material.

As can be seen in Fig. \ref{fig:The-return-rate} the return rate
of the most bound material happens on timescales comparable with scaling
for the low-mass ratio case (Eq.\ 1), but the profile of the return
rate is more steep, approximately $\sim t^{-7/3}$, with only a weak
dependence on the pericenter separation $R_{{\rm p}}$ for the cases
considered. We also note that in these simulations the fraction of
material bound to the black hole after the disruption is of the order
of 0.04 for a closest approach at the tidal radius ($R_{p}=2.15$
R$_{\odot}$), whereas impacts at deeper penetrations result in higher
fractions (0.06 and 0.2 for $R_{p}=1.94$ R$_{\odot}$ and $R_{p}=1.29$
R$_{\odot}$, respectively); for the following we adapt $f_{fall}=0.1$
as a typical value.

\begin{table}
\begin{tabular}{|c|c|c|c|c|}
\hline 
 &  & \multicolumn{3}{c|}{Mass Fraction}\tabularnewline
 & $R_{p}$ ($R_{\odot})$  & bound  & unbound  & left in star/planet\tabularnewline
\hline 
\hline 
Sun-like  & 1.29  & 0.2  & 0.1  & 0.7 \tabularnewline
\hline 
star  & 1.94  & 0.06  & 0.02  & 0.92 \tabularnewline
\hline 
 & 2.15  & 0.04  & 0.01  & 0.95 \tabularnewline
\hline 
Jupiter-like  & 1.29  & 0.49  & 0.49  & 0.02 \tabularnewline
\hline 
planet  & 1.94  & 0.49  & 0.49  & 0.02 \tabularnewline
\hline 
 & 2.15  & 0.49  & 049  & 0.02 \tabularnewline
\hline 
\end{tabular}

\caption{}
The fraction of bound fallback material and the return rate obtained
from hydrodynamical simulations (using the \textit{StarSmasher} code)
of tidal disruptions of a solar mass star and a Jupiter-mass planet
by a 10 M$_{\odot}$ BH. 
\end{table}

\subsubsection{Accretion time }

After circularization the bound debris from the tidal disruption forms
a torus or disk around the CO \citep[for example, see][]{Ruf92}
at radius of $r_{c}\simeq2R_{p}$. The accretion time scale for this
disk, assuming it is thick (with the ratio of disk scale height to
its radius, $h$, of the order of 1) is of the order of the viscous
time 
\begin{multline}
t_{{\rm acc}}\approx\frac{t_{{\rm kep}}(2R_{p})}{\alpha2\pi h^{2}}\\
\approx4.5\times10^{4}\,h^{-2}(\frac{\alpha}{0.1})^{-1}(\frac{R_{p}}{2.15R_{\odot}})^{3/2}(\frac{M_{\bullet}}{10M_{\odot}})^{-1/2}\,s\label{eq:accretion_time}
\end{multline}
where $t_{{\rm kep}}(r)$ is the Keplerian orbital period at orbital
radius $r$, $\alpha$ is the viscous constant. The value of the viscosity
constant $\alpha$ is unknown, and we normalize it with the commonly
used $\alpha=0.1$ value.

We therefore expect the formation of an accretion disk during the
few $1000$ s after the tidal disruption (Eq. \ref{eq:disruption_time}),
for the assumed parameters. This might be observable in the flare
rise time-scale. Later on the debris material will be accreted on
timescales of a few $10^{4}-10^{5}$ s onto the CO (Eq. \ref{eq:accretion_time}),
where as additional fall back material is expected to continuously
accrete onto the CO at a power law decaying rate (Eq. \ref{eq:debris_fall_back}
or somewhat steeper for the high-mass ratio disruption). The exact
timescales for the formation of the accretion disk is therefore less
important as the evolution is dominated by the longer accretion timescales.
We do note, however, that at the longer-term evolution as well as
in the case of the disruption of larger, evolved star (e.g.\ red
giant), occurring at a much larger radius, the accretion rate will
be dominated by the fallback time rather than the accretion timescale.
It was suggested that such a scenario may explain the origin of some
X-ray binaries (J. Steiner and J. Guillochon, private communication
2015; see also \citealt{1976MNRAS.175P...1H} and \citealp{kro84}
for tidal-capture-formed long-lived X-ray sources).

\subsubsection{Light curve}

Predicting the exact light curve due to the
accretion onto the compact object is difficult given the many uncertainties
in the evolution of the accretion-disk and the jet formation an evolution.
Nevertheless, in the following we speculate on the possible evolution
of the light curve, assuming a proportional relation between the material
accreted to the compact object and the luminosity produced by the
jet. 

\begin{figure}
\includegraphics[scale=0.4]{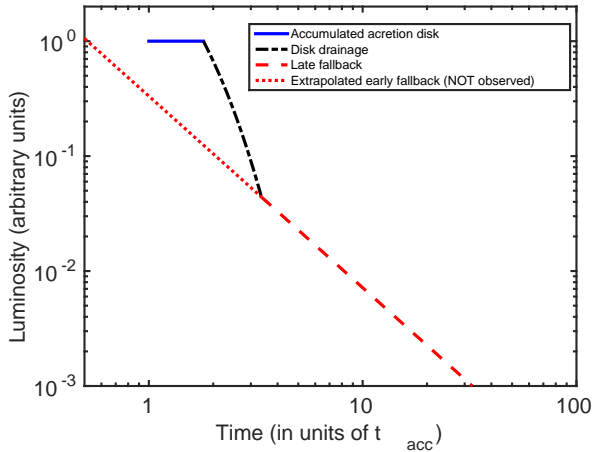}

\caption{\label{fig:light-curve}Schematics of the
expected $\mu$TDE light curve in the regime of $t_{min}\ll t_{acc}$
(see text), assuming the disruption occurred at time 0 (the flaring
taking place only after $\sim t_{acc}$).}
\end{figure}

We can generally divide the the $\mu$TDEs
into two regimes; (1) $t_{min}\gg t_{acc}$ and (2) $t_{min}\ll t_{acc}.$ 
\begin{itemize}
\item  $t_{min}\gg t_{acc}$: In this case, typically
corresponding to planet disruption (the low-mass ratio regime), the
fallback time scale is longer than the accretion time, and the accretion
evolution is therefore dominated by the fallback rate, i.e. it should
generally follow the regular $t^{-5/3}$ power-law, as derived for
TDEs by MBHs. 

\item  $t_{min}\ll t_{acc}$: For this case we can
summarize the results from the high mass-ratio disruption. We expect
the fallback material to accumulate and form a disk on the fallback
time, which then drains on the longer viscous time, and then keeps
a low accretion rate from the continuous tail of the fallback debris.
We would expect the flaring should therefore begin only once the material
is accreted on the compact object itself. We may therefore expect
four stages in the light curve evolution (see schematic representation
in Fig. \ref{fig:light-curve}): (1) A fast rise of the accretion
flare, once the disk material is processed in the disk and evolves
to accrete on the compact object; (2) Accretion from a disk until
the accumulated early fallback material is drained; again, we caution
that it's difficult to predict the exact shape of the light curve
at this stage, and it depends on the disk evolution and jet formation
mechanism; if we \emph{assume}
a steady state accretion until drainage, one might expect a relatively
 flat light curve. (3) Once the disk drains of the accumulated early
 fall-back material the light curve should drop steeply; (4) The continuous
 fall-back of material at the late stages would then govern the accretion
 rate at times longer than the viscous time, and the accretion rate
 should then follow the $t^{-5/3}$ rate (or somewhat steeper as discussed
 above). The exact light curve at the early stages is difficult to
 predict, but we do note a non-trivial signature of the $\mu$TDEs
 relating the early and late stages. Since the early accretion can
 only occur after the peak debris fallback, the \emph{extrapolation} of the late power-law ($t^{-5/3}$ or steeper) to early times (dashed
line; we emphasize that this extrapolation is \emph{not}
 the observed light curve), before the initial flaring, should always
precede the observed initial rise (see figure).
\end{itemize}

\subsection{Flare energy and spectra}

The flare energy corresponds to the accreted mass $E_{f}=\eta f_{acc}M_{\star}c^{2}$,
with $\eta$ the efficiency of transferring the rest mass to radiation
energy in the accretion process, and $f_{acc}$ the fraction of the
star mass which is accreted.

These accretion and flaring (e.g.\ through a jet production) processes
are still poorly understood, and the amount of accreted mass is strongly
dependent on the accretion scenario. Taking two very different scenarios
we try to give some possible lower and upper estimates on the radiated
energy released in this stage. We stress that the current lack of
knowledge on these accretion processes make more accurate calculations
suggestive at most.

If a fraction $\zeta$ of the circularized debris is accreted by the
CO, that is $f_{acc}=\zeta f_{fall}$, then 
\[
E_{f}=1.1\times10^{52}\,(f_{fall}/0.1)(\zeta/1)(\eta/0.1)(M_{\star}/0.6M_{\odot})\,ergs,
\]
which is somewhat lower but still comparable (especially for the disruption
of higher mass stars) with the isotropic equivalent energy of ultra-long
GRBs. The true released energy in observed GRBs should be reduced
as $E\Omega/4\pi$, with $\Omega$ the unknown GRB jet solid angle.
The derived $\mu$TDE energy could therefore be larger than required
for a GRB by orders of magnitude, if $\Omega\ll1$, and therefore
even much lower $f_{acc}$ could be sufficient. Such hyper-accretion
rates of $\dot{M}_{acc}\sim10^{10}\dot{M}_{Edd}$ are very similar
to that of an accreting CO in a common envelope with a massive star,
which have been studied both analytically (\citealt{1993ApJ...411L..33C,2000ApJ...541..918B}
and references within) and in simulations (e.g.\ \citealp{2000ApJ...532..540A}
and references within). Such accretion disks may be advection dominated
and quite possibly give rise to strong outflows and jets (see \citealp{2005Sci...307...77N}
for a short review).

\citet{2001ApJ...557..949N} suggest a convection dominated accretion
flow model to describe an accretion scenario, in which case only a
small fraction of the material is accreted, with a strong dependence
on the accretion disk outer radius and an appropriate decrease in
the flare energy. In this model, the accreted mass fraction is 
\begin{align*}
f_{acc} & \sim14.1f_{{\rm fall}}\left(\frac{\alpha}{0.1}\right)^{-1}\left(\frac{R_{{\rm out}}}{r_{{\rm s}}}\right)^{-1}\approx\\
\approx & 1.4\times10^{-4}f_{{\rm fall}}\left(\frac{\alpha}{0.1}\right)^{-1}\left(\frac{R_{p}}{2.15R_{\odot}}\right)^{-1}\left(\frac{M_{\bullet}}{10M_{\odot}}\right),
\end{align*}
where we have set $R_{{\rm out}}\approx2R_{p}$ for the outer radius
of the accretion disk and where $r_{{\rm s}}$ is the Schwarzschild
radius of the CO. In fact, a non-negligible fraction of the energy
released may not even be emitted but advected to the CO, at least
in the case of an BH accretor, which would make even this estimate
only an approximate upper limit for the flare energy in this case.

We conclude that the estimates for the energy emitted in the accretion
flare could vary by orders of magnitudes. If most of the tidal debris
mass is accreted to the CO, and the energy is efficiently emitted,
then such accretion flares may be as energetic as GRBs and should
be observable from extragalactic distances, though these would be
much fainter than typical GRBs due to their much longer duration.
Such flares could have GRB like characteristics but would not be necessarily
associated with a supernova. Indeed, \citet{luy+08} have suggested
such a related scenario for GRB 060614 where no associated supernova
was observed. They suggested a jet formation process scenario, which
could produce such a GRB following the tidal disruption of a solar
mass star by an intermediate massive black hole. Alternatively, as
different accretion scenarios suggest, large outflows from the accretion
disk may allow only a small fraction of the tidal debris to be accreted,
in which case much fainter flares will be produced.

We should stress that $\mu$TDEs are super-Eddington
events, and likely the radiation process arises from a jet. The exact
spectra from such jets and their formation are little understood,
but generally we would expect the flares to have non-thermal spectra
(unless significant interaction of the jet with the tidal debris thermalize
the emitted radiation, see e.g. suggestion by \citealt{Tho+11}).

\section{$\mu$TDE rates}

For a $\mu$TDE to occur we require a star/planet to pass by a CO
at a distance of no more than the tidal-disruption radius. In the
following we consider four possible scenarios in which such close
encounters can happen, and assess the $\mu$TDE rates and the expected
typical environment in which they occur in each of these cases. These
scenarios include (1) A CO has a random close encounter with another
star/planet in a dense stellar cluster. (2) A CO in a binary/planetary
system is kicked (e.g.\ through a NS/BH natal kick) and encounters
its close binary/planetary companion. (3) very wide binaries in the
field are lead into highly eccentric orbits due to multiple-scattering
with field stars, resulting in a close encounter.

Another possibility of encounter is through the secular evolution
of a CO-hosting quasi-stable triple system \citep{ant+12} leading
to close encounters (similar to the WD-WD collsions suggested by \citealp{kat+12}).
However, the uncertainties involved in this scenario, and in particular
the fraction of progenitor triple systems, are large, and the discussion
of this scenario is beyond the scope of this work.

\begin{table}
\begin{tabular}{|c|c|c|}
\hline 
Formation channel  & \multicolumn{2}{c|}{$\mu$TDE Rate {[}yr$^{-1}$ per MW-galaxy{]}}\tabularnewline
\hline 
 & BH  & NS\tabularnewline
\hline 
\hline 
Encounters in dense clusters  & $2.8\times10^{-6}$  & $4.8\times10^{-7}$\tabularnewline
\hline 
Natal kicks  & $1.4\times10^{-6}$  & $3.4\times10^{-7}$\tabularnewline
\hline 
Perturbed wide binaries  & $10^{-7}$  & --\tabularnewline
\hline 
\end{tabular}\caption{The rates of $\mu$TDEs from various potential channels. }
Note that the rate from the natal kick scenario corresponds only to
binaries with $>10$ AU separations. The perturbed wide-binaries channel
rates correspond to the high rates scenario obtained by \citet{mic+15},
for direct collapse BHs with no natal kicks (reproducing the LMXB
formation rate in the Galaxy); therefore this channel and the natal
kicks scenario are mutually exclusive. 
\end{table}

\subsection{$\mu$TDEs from stellar encounters in dense stellar clusters }

There are several scenarios in which a close encounter between a compact
object and a star/planet could occur. An isolated compact-object may
directly interact with a star or a binary in a two or three body encounter,
respectively. Similarly a CO in a binary system may interact with
another star or a binary system in three or four body interactions,
with encounters between higher-multiplicities possible. The dominant
type of encounter would depend on the environment and the multiplicity
of the systems \citep[and references therein]{lei+11}. In the following
we estimate the $\mu$TDE rate from such random encounters.

\subsubsection{Two body interactions}

The encounter rates between stars have been studied by many (see e.g.\ \citealp{1992ApJ...396..587D}
for very similar calculations). Here we give the rates of encounters
leading to a TDE, i.e.\ where the closest approach of a star/planet
to a given CO is smaller that the tidal radius. Such encounter rates
are dominated by the contribution from the densest stellar systems,
such as globular clusters (GCs) and galactic nuclei.

Consider a CO of mass $M_{co}$ in the core of a cluster containing
$N_{*}$ single stars (for a discussion of binaries, see below). A
tidal disruption will take place if the distance of closest approach
between this CO and a star is less than the tidal radius $R_{{\rm t}}$.
If all the ordinary stars had a mass $M_{*}$, the tidal disruption
rate from the one CO would be 
\[
\dot{p}_{co}(R_{t},\sigma)=2\pi G(M_{co}+M_{*})N_{*}R_{t}\sigma^{-1}V_{c}^{-1}
\]
where $\sigma$ is the relative velocity dispersion and the core volume
$V_{{\rm c}}$ can be written in terms of the core radius $R_{{\rm c}}$
of the GC as $V_{{\rm c}}=4\pi R_{{\rm c}}^{3}/3$. Here we are assuming
that the collision cross section is dominated by gravitational focusing.
The total disruption rate in a single GC is then 
\[
\Gamma_{co}\approx\dot{p}_{co}N_{co}\approx\dot{p}_{co}n_{0}f_{co}V_{{\rm c}},
\]
where $N_{co}$ is the total number of COs in the GC core, $n_{0}$
is the number density of stars in the core, and $f_{co}$ is the fraction
of COs in the stellar population (see \citealp{1992ApJ...396..587D}).

\noindent Typical GC cores have densities of $n_{0}\simeq10^{5}\:{\rm pc}^{-3}$,
and a typical core radius $R_{c}=1\:{\rm pc}$ \citep{pry+93,gne+97}.
Typical velocity dispersions in GC cores are roughly $\sigma\simeq20\:{\rm km\:s}^{-1}$
(see e.g. \citealt{1992ApJ...396..587D}). The fraction of COs in
the stellar population is taken to be $f_{{\rm NS}}=0.017$ and $f_{{\rm BH}}=0.012$
(taking a Salpeter mass function between $0.6-120$ ${\rm M_{\odot}}$
; NSs are assumed to originate from stars of mass between $8-15$
${\rm M_{\odot}}$ and BHs are assume to originate from stars more
massive than $15$ ${\rm M_{\odot}}$); however, due to NS natal kicks
and binary heating of BHs in cluster, only a fraction of these COs
are retained in the cluster. We assume a retention fraction of 0.05,
following \citet{2002ApJ...573..283P}, and we take typical neutron
star, and black hole masses to be, $1.4M_{\odot}$ and $10M_{\odot}$,
respectively. For COs of 1-10 $M_{\odot}$, the tidal radius falls
approximately in the range $R_{\odot}\lesssim r_{t}\lesssim2\,R_{\odot}$
for stars (and also for Jupiter like planets, whereas terrestrial
planets are disrupted only at $\sim2$ times smaller distances due
to their higher average densities). Using these typical values, we
calculate the disruption rates of stars by COs. We find $\Gamma_{{\rm NS}}=3.2\times10^{-9}\,{\rm yr}^{-1},$
and $\Gamma_{{\rm BH}}=1.8\times10^{-8}\,{\rm yr}^{-1}$. These rates
are consistent with more detailed cluster simulations where physical
collisions were considered (e.g. \citealp{iva+08}). Although the
population of BHs is smaller than that of NSs, their larger mass make
the encounter cross section much larger, thus enhancing the rate of
close encounters. For the $\sim$150 GCs observed in the MW galaxy
we finally get the tidal disruption rates per Milky-Way like galaxy
to be $\Gamma_{{\rm NS}}^{{\rm gal}}=4.8\times10^{-7}(f_{NS}/0.017)(f_{ret}/0.05)\,{\rm yr}^{-1},$
and $\Gamma_{{\rm BH}}^{{\rm gal}}=2.8\times10^{-6}(f_{BH}/0.012)(f_{ret}/0.05)\,\,{\rm yr}^{-1}$.
We point out that many galaxies contain much larger number of GCs,
and thus these estimates are only lower limits.

The rates of $\mu$TDEs of planets depend on the unknown fraction
of free floating planets in GC; for free floating planets to stars
ratio, $f_{ffp-star}$ of one (one free floating planet per star)
the rate should be slightly lower (due to the smaller combined mass),
but comparable to that of stellar disruptions, and the rate should
linearly scale with $f_{ffp-star}$.

\subsubsection{Three and four body interactions}

Three and four (or more) body interactions are much more complicated,
as they may involve resonant encounters in which the stars can pass
by each other several times. These could show chaotic behavior in
which the stars may pass each other at almost any arbitrary distance
\citep[see e.g. ][and references within]{2006tbp..book.....V}. Such
behavior can much enhance the possibility for very close passages
followed by tidal disruptions. Calculation of these rates require
better knowledge on the characteristics and distributions of binaries
and compact binaries in clusters, and a detailed treatment which is
beyond the scope of this work. However, from comparison with the somewhat
similar scenario of stellar collisions and tidal captures \citep{kro+84,2004MNRAS.352....1F,lei+11},
in which such encounters were found to have important contribution,
we note that our results thus give only a lower limit on the tidal
disruption rates, which could be higher by a factor of a few as a
result of these few body encounters.

\subsection{$\mu$TDEs from natal kicks of COs}

\label{sec:post kick disruption}

In the previous section we considered random encounters between a
star and an unrelated CO. Such an encounter can happen with a non-negligible
rates only in dense stellar clusters. A very different scenario for
encounters may arise in systems containing NSs or BHs with a stellar
or sub-stellar companion. In such systems the two companions may closely
interact following the natal kick imparted to the NS/BH at birth.

NSs and BHs are usually born following a violent supernova explosion
or through the coalescence of two COs (WD+WD, WD+NS, NS+NS). NSs,
and possibly BHs, are thought to be born out from these supernova
explosions with high velocities of tens to hundreds kilometers per
second (so called natal kicks, see e.g \citealp{2002ApJ...573..283P}
for a review and references). The comparison between the observed
high velocities of pulsars and the measured low velocities of their
progenitor stars are a strong indication for such kicks. BH formation
may also involve an intermediate stage of collapse into a NS, suggesting
BHs may acquire similarly high momentum kicks, leading to kick velocities
which are lower than the typical NS natal kick velocities, due to
the higher mass of BHs: $v_{{\rm kick}}^{{\rm BH}}=(M_{{\rm NS}}/M_{{\rm BH}})v_{{\rm kick}}^{{\rm NS}}$.
BHs formed through coalescence of COs might also acquire such high
velocities \citep{Ros+00}.

Following these high velocity kicks most binary systems would break
up, ejecting the newly formed NS/BH, and leaving behind their now
isolated companions. However, in some systems the kick imparted to
the newly formed NS/BH will give rise to a a close-approach trajectory
near the stellar/planetary companion (see e.g.\ \citealp{1994ApJ...423L..19L,2005MNRAS.361..955B}).
If the encounter is sufficiently close, the companion might be disrupted.
The total rates of $\mu$TDEs from such a scenario depend on many
uncertain parameters such as the fraction and semi major axis distribution
of such binary systems and the distribution of the velocity kicks;
a detailed population synthesis model to better evaluate the $\mu$TDE
rates from this scenario will be discussed elsewhere. In the following
we provide an order of magnitude estimate.

If the natal kick imparted to the CO after SN explosion is randomly
oriented and the kick velocity is larger than the Keplerian velocity
of the binary (i.e. the interaction is dominated by the natal kick
velocity), the probability a tidal disruption is the angular phase
space covered by the stellar companion target; i.e. the disruption
probability is of the order of $(\sigma_{t}/4\pi a^{2})$, where $\sigma_{t}=\pi R_{t}^{2}(1+2GM_{bin}/R_{t}v^{2}$)
is the cross section for such a close encounter (including the gravitational
focusing term), $v$ is the relative velocity, and $a$ is the binary
separation just after the SN. The probability for such events is therefore
a decreasing function of the binary separation.

Binaries with a massive primary ($>$ 8 M$_{\odot}$ progenitors of
NSs or BHs) and separations smaller than $\sim10$ AU will interact
through a common envelope. Let us first consider only binaries with
larger separations. Moe (PhD thesis, 2015 and private communication;
see also \citealp{san+12}) finds that $\sim80$ \% of all massive
stars have a binary companion with separation in the range $0.3-20$
AU, distributed with a log-uniform distribution. We therefore obtain
that $\sim0.3$ of all massive stars have non - strongly-interacting
binary companions in the separation range $10\,{\rm AU<r<20\,{\rm AU}}$.
Given the disruption probability dependence on the binary separation;
the rate of $\mu$TDEs will be dominated by binaries in this separation
range, and we neglect the additional smaller contribution from wider
binaries (only $\lesssim20$ \% of all massive stars have binary companions
with $a>20$ AU). The disruption probability is therefore on the order
of 
\begin{align*}
p & \sim\left(\frac{\sigma_{t}}{4\pi a^{2}}\right)\simeq\left(\frac{r_{t}^{2}}{a^{2}}\right)\left(\frac{GM_{bin}}{2r_{t}v_{kick}^{2}}\right)\simeq\\
\simeq & 5\times10^{-6}\left(\frac{M_{\bullet}/10M_{\odot}}{M_{\star}/M_{\odot}}\right)^{1/3}\left(\frac{R_{*}}{R_{\odot}}\right)\times\\
\times & \left(\frac{(M_{\bullet}+M_{\star})/11M_{\odot}}{(v_{kick}/190\,{\rm km\,s}^{-1})^{2}}\right)\left(\frac{15\,{\rm AU}}{a}\right)^{2}.
\end{align*}

The core-collapse SN rate in the Galaxy is $\sim2.8\pm0.6$ per century
\citep{li+11}. Therefore, assuming a binary fraction of $0.3$ in
the relevant separation range and $v_{kick}^{NS}\sim190\,{\rm km\,s}^{-1}$,
and $2/3$ of these SNe producing a NS, while the rest produce a BH,
we obtain a $\mu$TDE rate of $\sim3.4\times10^{-7}$ yr$^{-1}$ per
MW galaxy for NSs, while the more massive (and hence smaller kick
velocity) BHs provide a rate of $\sim1.4\times10^{-6}$ yr$^{-1}$
per MW galaxy.

Interacting binaries which evolve through mass transfer may go through
a common envelope and either merge or give rise to a compact binary
with a short period, prior to the SN explosion. In such compact binaries
the orbital Keplerian velocity is high and could be comparable with
or much higher than the kick velocity, such that the conditions for
a close approach following the natal kick need to be fine tuned \citep{Tro+10}.
\citet{Tro+10} studied in detail the possibility of a collision between
two COs in a close binary following a SN kick, obtaining collision
probabilities of the order of $10^{-8}-10^{-7}$. The cross-section
for a tidal encounter with a star is much higher than the cross-section
for a collision (by a factor of the tidal radius to the CO radius
in the gravitational focusing regime), suggesting this channel as
a potentially promising channel for $\mu$TDEs, with tidal disruption
probability of the order of $10^{-4}-10^{-3}$. Nevertheless, the
formation rate of such short period binaries just prior to the SN
explosion strongly depends on not well understood binary stellar evolution
and in particular the common envelope phase. We do note that X-ray
binaries evolve through a short-period binary phase, and that their
formation efficiency is low (the formation rate of X-ray binaries
is of the order $\sim3\times10^{-7}$ yr$^{-1}$ per MW galaxy) suggesting
that this channel might not be very efficient. 

\subsection{$\mu$TDEs from perturbed wide-binaries in the field }

\citet{mic+15} suggest that low-mass X-ray binaries might form in
the field through close interactions between a CO and a wide binary
companion induced by perturbations from field stars (following \citealp{kai+14},
who discussed a similar scenario for the collisions of two MS stars).
Such a process leads to tidal capture like processes, and may therefore
produce tidal disruption events at similar rates. Michaely \& Perets
consider various scenarios; the most efficient scenarios can reproduce
the LMXB population in the Galaxy. If we assume this is indeed the
main channel for LMXB formation (especially BH-LMXBs) we should expect
a similar rate of $\mu$TDEs from the same scenario, namely of the
order of few times $10^{-7}$yr$^{-1}$ per MW galaxy.

\section{Discussion and Summary }

In this paper we have explored the partial and full tidal disruptions
of stellar and sub-stellar objects by stellar COs. Such disruptions
may result in energetic flares of long duration (tens of minutes to
hours) following the accretion of the tidal debris onto the CO. The
flare energy is highly dependent on the poorly understood accretion
process, and could vary by orders of magnitude. If most of the tidal
debris mass is accreted to the CO, and the energy is efficiently emitted,
the accretion flares may be as energetic as GRBs and should be observable
from extragalactic distances, probably with GRB like characteristics
but with much longer timescales ($10^{3}-10^{4}$ s; and hence fainter),
starting with a relatively flat light curve
but decaying in a power law fashion at late times. Such flares are
also not necessarily associated with a supernova (where GRB 060614
may serve as a possible candidate; see \citealt{luy+08} for a similar
suggestion). Alternatively, large outflows may allow only a small
fraction of the tidal debris to be accreted, in which case much fainter
flares, with total energies smaller by orders of magnitudes may arise.

Our main focus was on disruptions by BHs. Though many similar aspects
are expected to characterize disruptions by NSs and WDs, the latter
have a physical surface and the accreted material may interact with
the surface possibly producing violent events such as X-ray bursts
(NSs) and novaes (WDs). Moreover, the accretion of material through
TDEs may affect the evolution of such COs and their spin, in cases
where significant amount of material is accreted. This work focuses
on the accretion event itself, but future follow-up should address
the long-term implications and observational signatures from $\mu$TDEs,
which are beyond the scope of our current discussion.

On the longer timescales debris from a $\mu$TDE may slowly continue
to fall back, possibly forming a long-lived accretion disk (e.g.\ especially
if the disrupted star is an evolved star) that could power an X-ray
binary-like object, though with only an accretion disk and not an
actual stellar companion (J.\ Steiner and J.\ Guillochon, private
communication 2015; see also \citealt{1976MNRAS.175P...1H,kro84}
for a consideration of tidal-capture-formed long-lived X-ray sources).
In an alternative scenario the debris that falls back on the CO gives
rise to a gaseous envelope around the CO, possibly forming a Thorne-Zytkow
object; however this requires the fall-back of the disrupted star
to be very large \citep{1977ApJ...212..832T}, likely arising only
from a direct collision rather than a tidal disruption.

Flares following close encounters in dense stellar systems, are likely
to occur mainly in dense systems such as globular clusters, galactic
nuclei or massive young clusters, and may thus be observable in both
early and late type galaxies, with total $\mu$TDE rates of $10^{-7}-10^{-6}$
yr$^{-1}$ per MW-galaxy; comparable rates might also be obtained
through tidal disruptions occurring in perturbed wide (>1000 AU) binaries
in the field. Flares following a kick of the CO into stellar/sub-stellar
companion are likely to be associated with a SN and would occur in
star-forming regions, typically months after the SN. Such $\mu$TDEs
can also occur at rates comparable with the other scenarios, but are
rare compared to the core-collapse SNe rate, i.e. they occur once
for every $\sim10^{5}$ core-collapse SNe. We do caution, however,
that the latter rate may be underestimated since it does not include
the potential contribution from kicks in short-period planetary systems
or binaries (possibly evolved through a common-envelope phase prior
to the SN; $\mu$TDEs in the latter systems may occur hours to days
after the explosion (i.e.\ during the early stages of the SN rise).
The ultra-long GRB 111209A has been observed to be associated with
a very luminous SN \citep{Gre+15}. The delay between the SN and the
GRB in this case was at most a few days, and therefore, a $\mu$TDE
interpretation for this event indeed requires a compact, likely post-common
envelope binary. We should stress that $\mu$TDEs are super-Eddington
events, and likely the radiation process arises from a jet. In this
case he \emph{observed} rates should be reduced compared with our
calculations by a beaming factor, while the apparent luminosities
we describe should be enhanced by the same factor.

The expected properties of $\mu$TDEs are
consistent with and might explain the origins of ultra-long GRBs \citep{lev+14};
long lived ($10^{4}$ s) GRBs, showing an initial plateau followed
by a rapid decay. Our current models suggest that $\mu$TDEs are not
likely to explain the origin of
GRBs. 

$\mu$TDEs may also explain the origin of
the SWIFT detected TDE-candidates \citep{Blo+11,Bur+11,Can+12,Bro+15},
suggested to be produced through a TDE by a supermassive black hole.
The typical timescales for the latter are longer than the observed
$10^{5}$s, challenging the currently suggested origin, but quite
consistent with the $\mu$TDE scenario. 

The possible cases where $\mu$TDEs could be related to regular GRBs
are those in which a GRB results in the formation of a CO, which is
then kicked and disrupts a companion. In this respect, we can mention
the very long flare (a $few\times10^{4}\,s$ ) observed in GRB 050724
a few hours after the prompt emission, which could possibly be explained
by a $\mu$TDE. Interestingly, this would come in accord with a scenario
suggested by \citet{2005astro.ph.10192M} for this GRB. They try to
explain a shorter late flare in this same event by shock heating from
the prompt GRB explosion on a companion which they suggest exists
for this GRB progenitor. Taking into account the appropriate velocities
possible for a kick, the timescale for both events (the shock heating
and the tidal disruption) would correspond to the same distance between
the binary members. However, the relevant timescales as well as other
flares observed in this event make other scenarios equally, if not
more, plausible than a disruption event.

The late flare (16 days after the GRB) in the case of GRB 050709 \citep{2005Natur.437..845F}
is also noteworthy. The very long delay between the flares, could
potentially be explained by a $\mu$TDE event; e.g.\ produced following
a natal kick during the formation of a BH from the merger of two NSs,
which then disrupts a wider companion. Such a scenario might be fine
tuned, but currently no other scenario for this extremely late flare
had been suggested.

Finally, the recently discovered sample of ultra-long GRBs could potentially
be explained as $\mu$TDEs. Such a scenario would naturally explain
their very long timescales compared with regular long GRBs, and suggest
the possible existence of yet longer time-scale, but fainter events.

\acknowledgements{We thank T. Alexander, A. Loeb and Eli Waxman for helpful discussions.
H.B.P. acknowledges support the Israel Science Foundation excellence
center I-CORE grant 1829. ZL acknowledges support from the National
Natural Science Foundation of China (No. 11273005) and the National
Basic Research Program (973 Program) of China under grant No. 2014CB845800.
JCL and SRM are supported by the National Science Foundation (NSF)
grant number AST-1313091.}

\expandafter\ifx\csname natexlab\endcsname\relax\global\long\def\natexlab#1{#1}
 \fi

\bibliographystyle{aasjournal}

\end{document}